\documentclass{elsart}
\usepackage{graphicx}
\usepackage{amssymb}
\usepackage{upgreek}
\usepackage{nicefrac}
\usepackage{rotating}
\usepackage[english,francais]{babel}

\def\og{\leavevmode\raise.3ex\hbox{$\scriptscriptstyle\langle\!\langle$~}}
\def\fg{\leavevmode\raise.3ex\hbox{~$\!\scriptscriptstyle\,\rangle\!\rangle$}}

\renewcommand{\vec}[1]{\mathbf{#1}}

\begin{document}
% Select a primary header Physics or Astrophysics
% You can place after the header (classification), if you know it.

%\centerline{Physics or Astrophysics/Header}
\begin{frontmatter}


% \author{Name\thanksref{label2}}
\selectlanguage{english}
\title{Bismuth-based perovskites as multiferroics}

% If all authors are at the same address, the [label1] can be suppressed

\selectlanguage{english}
\author[list]{Mael Guennou},
\ead{mael.guennou@list.lu}
\author[cea]{Michel Viret}
\ead{michel.viret@cea.fr}
\author[list,uni]{Jens Kreisel}
\ead{jens.kreisel@list.lu}

\address[list]{Materials Research and Technology Department, Luxembourg Institute of Science and Technology, 41 rue du Brill, L-4422 Belvaux, Luxembourg}
\address[cea]{Service de Physique de l'\'Etat Condens\'e, CEA Saclay, DSM/IRAMIS/SPEC, UMR CNRS 3680, 91191 Gif-Sur-Yvette Cedex, France}
\address[uni]{Physics and Materials Science Research Unit, University of Luxembourg, 41 Rue du Brill, L-4422 Belvaux, Luxembourg}

% If your know the dates of reception, and acceptation you can put them now;
%    idem the name of the person presenting your article

\medskip
\begin{center}
{\small Received *****; accepted after revision +++++}
\end{center}

\begin{abstract}
This review devoted to multiferroic properties of Bismuth-based perovskites falls into two parts. The first part focuses on BiFeO$_3$ and summarizes the recent progress made in the studies of its pressure-temperature phase diagram and magnetoelectric coupling phenomena. The second part discusses in a more general way the issue of polar -- and multiferroic -- phases in Bi$B$O$_3$ perovskites and the competition between ferroelectricity and other structural instabilities, from an inventory of recently synthetized compounds. 

{\it To cite this article: M. Guennou, M. Viret and J. Kreisel, C. R. Physique ** (****).}

\vskip 0.5\baselineskip

\selectlanguage{francais}
\noindent{\bf R\'esum\'e}
\vskip 0.5\baselineskip
\noindent
{\bf Multiferro\"icit\'e dans les p\'erovskites au bismuth.}
Cette revue consacr\'ee aux p\'erovskites multiferro\"i{}ques \`a base de bismuth Bi$B$O$_3$ est scind\'ee en deux parties. La premi\`ere est consacr\'ee au cas de BiFeO$_3$ et r\'esume les progr\`es r\'ecents r\'ealis\'es dans l'\'etude de son diagramme de phases pression-temp\'erature, et de ses ph\'enom\`enes de couplage magn\'eto-\'electrique. La seconde partie aborde de mani\`ere plus g\'en\'erale, \`a partir d'un inventaire des compos\'es r\'ecemment synth\'etis\'es, la question de la stabilit\'e des phases polaires -- et multiferroiques -- dans les p\'erovskites Bi$B$O$_3$ et la comp\'etition entre la ferro\'electricit\'e et les autres instabilit\'es structurales.
  
{\it Pour citer cet article~: M. Guennou, M. Viret and J. Kreisel, C. R. Physique ** (****).}

%Now keywords/mots-clÈs
 \vskip 0.5\baselineskip
\noindent{\small{\it Mots-cl\'es~:} Multiferro\"\i{}que~; P\'erovskite~; BiFeO$_3$}}
\end{abstract}
\end{frontmatter}

% now the Version franÁaise abrÈgÈe, if it exists
%\selectlanguage{francais}
%\section*{Version fran\c{c}aise abr\'eg\'ee}
% Text of your Version franÁaise abrÈgÈe here

\selectlanguage{english}
% main text

\section{Introduction}

Multiferroics are materials that possess simultaneously magnetic and ferroelectric order. Although found separately in a large number of $AB$O$_3$ perovskites (along with virtually the full scope of possible functional properties including superconductivity, piezoelectricity, insulating/metal/semiconducting behavior etc.), the simultaneous combination of ferroelectricity and magnetism in a single phase $AB$O$_3$ perovskite is rather scarce. The reason for this was rationalized in a seminal paper by N. Hill, who pointed out that in most classical ferroelectrics, typically the titanates PbTiO$_3$ and BaTiO$_3$, the polar cation shifts are caused by a second-order Jahn-Teller effect, or hybridization, involving $B$-site ions with d$^0$ electrons, whereas in contrast, the presence of unsaturated $d$ electrons is required for the transition elements to acquire a magnetic moment. In order to waive this contradiction and reconcile ferroelectricity and magnetism, one therefore has to induce ferroelectricity by other mechanisms. One such alternative is found in rare-earth manganites at low temperatures, where ferroelectricity is induced by the cycloidal spin order; this has defined the so-called "Type II multiferroics" \cite{Khomskii2009}. Another track that has been followed in the recent years is to induce ferroelectricity and magnetism on two different crystallographic sites of the perovskite, namely bismuth on the $A$-site and a magnetic transition metal (Fe, Mn etc.) on the $B$-site. The ferroelectric off-centering of the Bi$^{3+}$ cation is favored by the 6s$^2$ electron "lone pair. This type of multiferroic is best exemplified in BiFeO$_3$ (BFO), which has become a model system for so-called "Type I" multiferroics and has stimulated a significant international research effort over the past years. 

In our short review we will in a first part discuss recent advances on BFO. For this, we will take a 2009 published review as a starting point \cite{Catalan2009} and focus on the progress made since then on the pressure-temperature phase diagram, the magnetoelectric coupling and exchange bias. In the second part, we shall move away from the special case of BFO and address the issue of ferroelectricity in Bi-based perovskites in general, in the light of the recent progress in synthesizing these compounds, especially with magnetic cations on the $B$-site, and subsequent studies. 

\section{BiFeO$_3$ updated}

\subsection{Structural and ferroelectric properties}

\begin{figure}
\begin{center}
\includegraphics[width=0.7\textwidth]{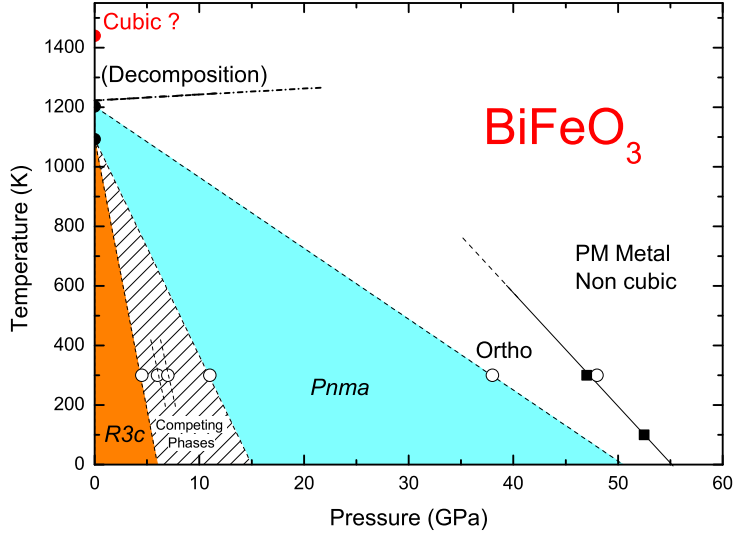}
\caption{Pressure-temperature diagram for bulk BFO, reproduced from Ref. \cite{Guennou2011a}. The transition temperature to a cubic phase at ambient pressure lies beyond the decomposition temperature and is a theoretical prediction \cite{Kornev2007a}.}
\label{fig1}
\end{center}
\end{figure}

The room-temperature structure of BFO, also called $\upalpha$ phase, is a distorted perovskite structure with a slight rhombohedral distortion \cite{Lebeugle2007}. It exhibits both large octahedra tilts along the $[111]$ pseudo-cubic direction and a strong ferroelectric polarization, both treated as pseudo-proper order-parameters. This is a major originality of the BFO structure as compared to classical ferroelectrics. It notably contrasts with the typical compounds CaTiO$_3$ and PbTiO$_3$, where in both cases, the cubic cell is calculated to be unstable upon both tilt and ferroelectric instabilities, but in practice, the crystal lowers its energy by developing one instability only (tilts for CaTiO$_3$, polar displacements for PbTiO$_3$) while the second one is inhibited. A consequence of these two interacting instabilities in BFO is a rather complex temperature-pressure phase diagram which is characterized by multiple phase transitions as illustrated in Fig. \ref{fig1}. On the theory side, this richness is perhaps best illustrated by the first-principles calculations by Di\'eguez et al. \cite{Dieguez2011}, who have found an unusually large number of (meta)stable structures for BFO with different ferroelastic, ferroelectric and magnetic properties, with the perspective of multiferroic properties and strong magnetoelectric coupling at room temperature. Recent reports of a structural phase transition close to room temperature in highly compressively strained BFO thin films add further support to this \cite{Kreisel2011,Siemons2011,Infante2011a}. The recent advances in thin film structures, also of considerable interest, are out of scope here but can be found in another chapter of this review by Yang {\it et al.} and in Ref. \cite{Sando2014}.

Experimentally, the phase sequence with temperature has been a long standing controversy. Two phase transitions from the $\upalpha$ to the so-called $\upbeta$ and $\upgamma$ phases have been identified some time ago at 825$^\circ$C (1098~K) and 931$^\circ$C (1204~K) before decomposition occurs just about 10$^\circ$C higher. The determination of the crystal structure of the $\upbeta$ and $\upgamma$ phases has been very difficult and controversial, due to the proximity of the decomposition temperature. The most recent neutron studies \cite{Arnold2009,Arnold2010} indicate that both the $\upbeta$ and $\upgamma$ phases are paraelectric orthorhombic phases with the $Pnma$ structure, the $\upalpha\rightarrow\upbeta$ then corresponding the Curie temperature $T_\mathrm C$ and the $\upbeta\rightarrow\upgamma$ transition being an isostructural transition, also associated to an insulator-to-metal (IM) transition.  

The phase transition sequence under high pressure is even more complex with, remarkably, six phase transitions in the 1~bar to 60~GPa range \cite{Guennou2011a}. This again contrasts with the classical ferroelectrics and ferroelastics, which display in the same pressure range one or two structural transitions \cite{Ishidate1997,Janolin2008,Guennou2010,Guennou2010a}. An orthorhombic phase with $Pnma$ symmetry is stable in a very large pressure range between 11 and 38~GPa, and the experiments suggest that the structure changes very little in terms of strain and tilt angle in this pressure range, and remains essentially identical to the $\upbeta$ phase. 

In the lower pressure range, four structural transitions at 4, 5, 7, and 11~GPa have been observed. In this range, BFO displays complex domain structures and unusually large unit cells, with lattice parameters given by $a\approx\sqrt 2a_\mathrm{pc}$, $b\approx 3\sqrt 2a_\mathrm{pc}$, $c\approx 2a_\mathrm{pc}$, which suggests a competition between complex tilt systems and possibly off-center cation displacements. Moreover, it has been shown that non-hydrostatic shear stress strongly affects the observed phase sequence \cite{Guennou2011}. The detailed structure of these phases remains unclear, but theoretical work have suggested the presence of "nanotwinned" structures, with complicated tilt patterns that go beyond the classical description by Glazer \cite{Prosandeev2013}. In those phases, the tilt patterns are not limited to being "in-phase" or "anti-phase" in adjacent layers, but can adopt more complex sequences with a longer period (e.g. 4 unit cells or more), giving rise to complicated structures with very large unit cells, similar to those observed experimentally. 

In the very high pressure range, BFO becomes again unstable and shows two further pressure-induced phase transitions at 38 and 48~GPa which are marked by the doubling of the unit cell and an increase of the total distortion. Resistivity measurements have revealed that the last transition around 48~GPa is also an insulator-to-metal (IM) transition \cite{Gavriliuk2005,Gavriliuk2007}. IM transitions are also observed in rare-earth orthoferrites in the 40-50 GPa range \cite{Rozenberg2005}, but follow a different pattern, with a very strong volume drop explained by a high-spin to low-spin transition and an isostructural $Pnma\rightarrow Pnma$ transition. In contrast, the transition in BFO is symmetry breaking and does not exhibit a significant volume drop \cite{Guennou2011a}.

One striking feature of the phase diagram as a whole is the absence of the simple cubic phase, both at high pressures and high temperatures. At high temperatures, the transition to the cubic phase has been predicted by ab-initio calculations above 1400~K \cite{Kornev2007a}, i.e. above the decomposition temperature, so that it is not accessible experimentally. This contrasts with lead-based ferroelectrics where the cubic phase is always reached at "reasonable" temperatures. Under high pressure, not only is the cubic phase absent, but, in addition, the total strain increases under pressure. In other words the crystal moves away from the cubic structure. This can be understood if the crystal structure becomes dominated by ferroelastic antiferrodistortive instabilities.  

Yet, the full phase diagram, being almost based on one isotherm and one isobar, remains hypothetical. There is no real justification for the straight lines connecting the observed transitions. In particular it is not at all clear whether or not the region between 4 and 11~GPa at ambient pressure, with many complex and competing phases, extends up to $T_\mathrm C$ at ambient pressure. But we note that, if it does, it would in part explain why the structure determination of the $\upbeta$ phase has been so problematic. Experimental investigations of the high-temperature high-pressure part of the phase diagram are desirable to clarify this. 

\subsection{Magnetic properties and magneto-electric coupling}

\subsubsection{Magnetic structure}

\begin{figure}
\begin{center}
\includegraphics[width=0.7\textwidth]{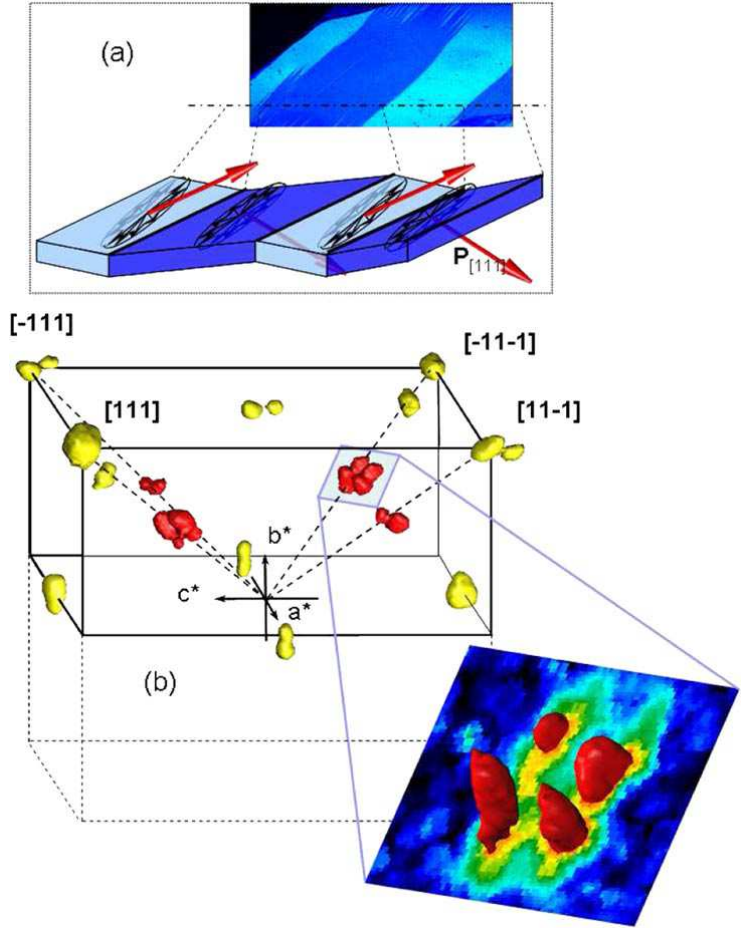}
\caption{Mapping of the neutron intensity in reciprocal space. The yellow spots are purely nuclear intensities and red spots are purely magnetic. Two sets of splitting appear for the nuclear intensity: one because of a difference in reticular plane distance due to the two rhombohedral distortions along $[111]$ and $[1\overline 11]$, and the second because of a physical buckling of the crystal induced by the twinning. Magnetic peaks are further split because of the cycloids. The patterns are consistent with the presence of two ferroelastic domains with $\vec P$ along $[111]$ and $[1\overline 11]$. Note that because the splitting is small, the scale has been magnified by a factor of 10 on each peak position. Reprinted from Ref. \cite{Lebeugle2008}.}
\label{fig2}
\end{center}
\end{figure}

The most powerful means of studying the magnetic structure of bulk samples has been neutron diffraction, first on powder samples and in more recent times on single crystals. Early powder diffraction measurements have established that BFO is a G-type antiferromagnet \cite{Sosnowska1982,Przenioslo2006}. The ordering of Fe$^{3+}$ ions is that of a nearest neighbour antiferromagnetic arrangement, a structure very easy to identify in neutron measurements because of the presence of peaks at the $(\pm\nicefrac{1}{2},\pm\nicefrac{1}{2},\pm\nicefrac{1}{2})$ vectors in reciprocal (cubic) space. Single crystal data can be more instructive in giving the exact magnetic configuration. Figure \ref{fig3} shows the 3D reciprocal space mapping of a crystal \cite{Lebeugle2008} where yellow peaks are of nuclear origin and the red peaks are magnetic. In that spectrum, the four nuclear peaks in the pseudo cubic diagonals, the $(111)$ and $(1\overline 11)$ reflections, are split along directions following the long diagonals (dashed lines on the figure). This indicates the presence of two rhombohedral twins consistent with the existence of two ferroelectric domains in the crystal with polarization axes along $(111)$ and $(1\overline 11)$. One can also notice that the other $(\overline 111)$ and $(11\overline 1)$ reflections are also split, but along the $(101)$ direction. This is due to the buckling of the crystal (schematically shown in the inset of fig. \ref{fig3}) which slightly changes the angles fulfilling the Bragg conditions. In this experiment, the neutron data were taken after the initially single ferroelectric state crystal was driven into a bidomain state by applying a voltage in the $(001)$ direction. This multi-domain state consists of stripe like regions with two different polarization directions which produce the double splitting of interest in the figure. 

\subsubsection{Magneto-electric coupling}

When magnetic ions also participate in the ferroelectric order, the magnetoelectric coupling is said to be direct. An electric field will act on the dipolar moments and slightly change the angle of the bonds between magnetic and oxygen ions thus changing the exchange integral responsible for the magnetic order. This is a direct effect induced by an applied, or internal, field. Indeed, internal fields are also responsible for a coupling energy responsible for a 'spontaneous magnetoelectric effect' \cite{Agyei1990}. In compounds where the electrical polarization is induced by the magnetic order, the two ordering temperatures coincide and the ferroelectric is called 'improper'. The polarization is generally at least 1000 times weaker than in a classical ferroelectric but the magnetoelectric coupling is expected to be strong. The other type of magnetoelectric coupling, said to be 'indirect', appears when magnetism and ferroelectricity are due to two different sublattices. The compounds can be either one or two intimately mixed materials. In these compounds, the magnetic ions will be displaced by magnetostriction and the interaction with the ferroelectric system is mediated by the elastic properties. 

\paragraph{Direct coupling of polarization and magnetic orders}

\begin{figure}
\begin{center}
\includegraphics[width=0.7\textwidth]{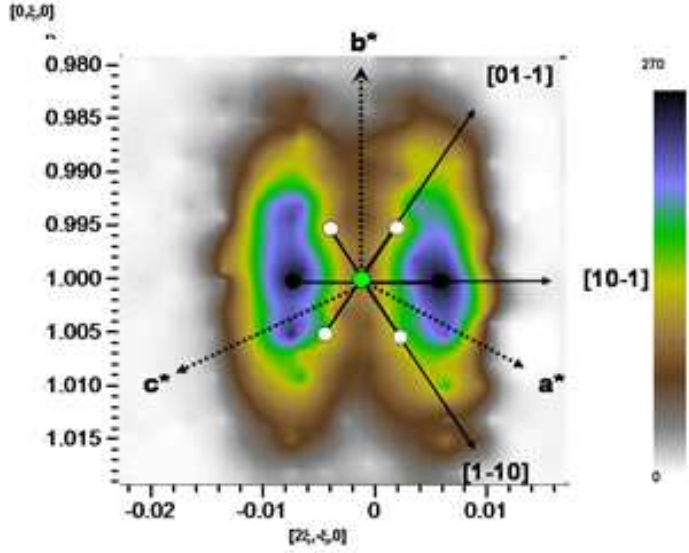}
\caption{Neutron intensity around the $(\nicefrac{1}{2},-\nicefrac{1}{2},\nicefrac{1}{2})$ Bragg reflection in the single domain state. The two diffraction satellites indicate that the cycloid is along the $(10\overline 1)$ direction. Reprinted from Ref. \cite{Lebeugle2008}.}
\label{fig3}
\end{center}
\end{figure}

\begin{figure}
\begin{center}
\includegraphics[width=0.5\textwidth]{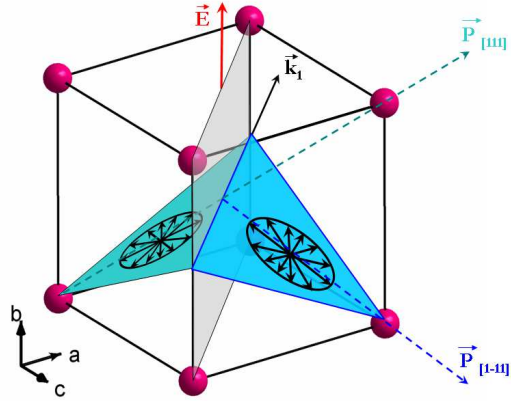}
\caption{Schematics of the planes of spin rotations and cycloids $\vec k$ vector for the two polarisation domains separated by a domain wall (in grey). Reprinted from Ref. \cite{Lebeugle2008}.}
\label{fig4}
\end{center}
\end{figure}

The purely antiferromagnetic peaks of $(\pm\nicefrac{1}{2},\pm\nicefrac{1}{2},\pm\nicefrac{1}{2})$ type (in red in Fig. \ref{fig2}) have been analyzed in more details \cite{Laukhin2006}. They are composed of four spots as shown in the zoomed region of Fig. \ref{fig3}. These can be understood by considering first the ferroelastic effects evidenced in the nuclear peaks: the intensity from $\vec P_{111}$ and $\vec P_{1\overline 11}$ domains is obtained at a slightly different angle. The other splitting in the $(\overline 101)$ direction corresponds to magnetic satellites due to periodic incommensurate arrangements already seen in older powder diffraction data \cite{Sosnowska1982}. In single crystals, the splitting indicates directly the direction and value of the magnetic structures' propagation vectors. The pattern is even simpler for the monodomain crystal where, before application of an electric field, Lebeugle et al. showed that there existed only a single incommensurate structure (fig. \ref{fig4}). Because of the presence of a threefold axis in the rhombohedral cell, there are indeed three symmetry-equivalent propagation vectors for this structure: $\vec k_1 = (\delta\ 0\ -\!\!\delta)$, $\vec k_2 = (0\ -\!\!\delta\ \delta)$ and $\vec k_3 = (-\delta\ \delta\ 0)$ where $\delta = 0.0225$. In neutron diffraction, these different $\vec k$ domains lead to a splitting of magnetic peaks along three different directions. In a powder sample, all three $\vec k$ domains are equally populated and the determination of the modulated magnetic ordering is not unique. Indeed, pure cycloids, elliptical cycloids and spin density waves, give the same diffraction pattern \cite{Przenioslo2006}. The exact nature of the periodic structure is an important parameter in antiferromagnetic ferroelectrics since recent models of the magnetoelectric coupling give non-vanishing electric polarization $\vec P$ for cycloids and elliptic ordering and zero polarization for a spin density wave \cite{Sampaio2003,Bea2008}. This ambiguity can be eliminated using a single crystal with a unique periodic structure as in Fig. \ref{fig3} where high-resolution scans around the strongest magnetic reflections of the $(\pm\nicefrac{1}{2},\pm\nicefrac{1}{2},\pm\nicefrac{1}{2})$ type provide the necessary information. The measurements of fig. \ref{fig3} demonstrate that the modulated structure has a unique propagation vector $\vec k_1= \pm(\delta\ 0\ -\!\!\delta)$ while scans around the other AF positions confirm that the crystal is in a single magnetic domain state. A fit of the intensity convoluted with the resolution function gives a modulation period of 64(2)~nm, in agreement with that reported for powder samples. Furthermore, the spin rotation plane can also be determined. A quantitative analysis \cite{Laukhin2006} of the integrated intensities of ten $\theta/2\theta$ magnetic reflections allows to conclude unambiguously that the structure is a circular cycloid with antiferromagnetic moments $\upmu$(Fe) = 4.11(15)$\upmu_B$ lying in the plane defined by $\vec k_1=(\delta\ 0\ -\!\!\delta)$ and the polarization vector $\vec P = (111)$. A consequence of the single $\vec k$ vector of the cycloid is that the crystal symmetry is lowered. Indeed, the ternary axis is lost and the real average symmetry becomes monoclinic with the principal axis along $\vec k$ (i.e. $(110)$ directions).

After application of an electric field and the creation of the bi-domain state, the diffraction pattern of fig. \ref{fig2} indicates that, in this particular case, the cycloids in both domains kept their original propagation vectors. However, there are now two rotation planes of the AF vectors in the cycloids indicating that half of the crystal volume had switched its polarization by 71$^\circ$, and brought with it the plane of rotation of the Fe moments which toggled from the original $(\overline 12\overline 1)$ plane to $(121)$. The propagation vectors thus remained in the $(\overline 101)$ direction and both cycloids lay in the plane defined by $\vec k$ and $\vec P$. This effect corresponds to an electric field induced spin flop of the antiferromagnetic sublattice and constitutes a direct proof of the significant coupling between the polarization vector and the magnetic cycloid of the Fe$^{3+}$ moments. The resulting schematic magnetic and electric configuration of the two domains is represented in fig.~\ref{fig4}.
 
\begin{figure}
\begin{center}
\includegraphics[width=0.7\textwidth]{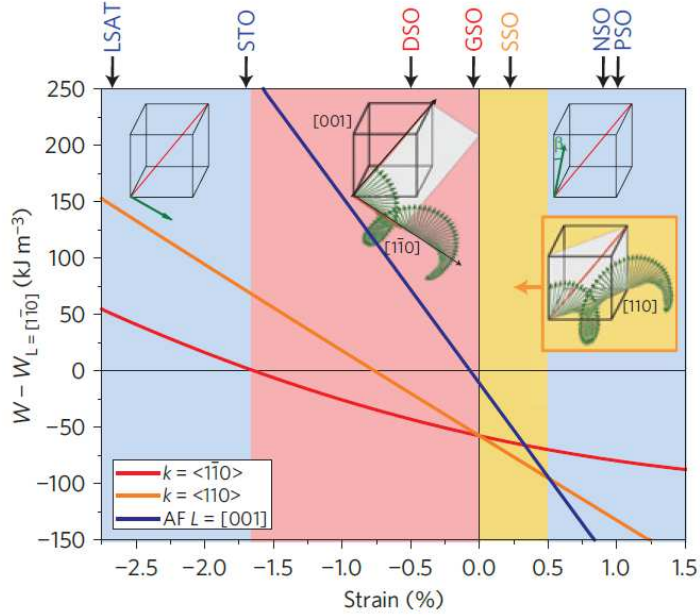}
\caption{The energy of three magnetic states (bulk-like "type-1" cycloid with propagation vector along $\langle 1\overline 10\rangle$ directions, "type-2" cycloid with propagation vector along $\langle 110\rangle$ directions, and collinear antiferromagnetic order with antiferromagnetic vector close to $[001]$), relative to a fourth magnetic state, corresponding to a collinear antiferromagnetic order with antiferromagnetic vector along in-plane $[1\overline 10]$ directions. The stability regions of the different states are shown in colours (blue: antiferromagnetic; red: type-1 cycloid; orange: type-2 cycloid). The different substrates used are located on top of the diagram at their corresponding strain. Reprinted from Ref. \cite{Sando2013}.}
\label{fig5}
\end{center}
\end{figure}

Beside these intrinsic properties of BFO, it is important to notice that thin films strained on a substrate can show rather different behavior. Even though ferroelectric properties do not seem to be very affected, the cycloids often completely disappear. Indeed, the presence of a strain induced anisotropy axis can destabilize the cycloidal arrangement. The magnetic phase diagram of BFO films deposited on a number of substrates imposing different strains has recently been established   as shown in fig. \ref{fig5}. While at high epitaxial strain the cycloidal modulation is destroyed, it is observed that non-collinear orders are stable at low strain. Interestingly, the cycloidal wavevector can change direction at intermediate strain states and spins progressively reorient from in-plane to out-of-plane as strain goes from compressive to tensile. The vast majority of published results on thin films are on SrTiO$_3$ substrates inducing a large tensile strain where the cycloid disappears leaving a simple G-type antiferromagnetic order with a slight global spin canting of 0.7$^\circ$ \cite{Bea2007}. Nevertheless, magnetoelectric coupling also exists as demonstrated using synchrotron PEEM \cite{Zhao2006a}. The coupling mechanism cannot be that due to the incommensurate magnetic structure and it is probably of magnetostriction origin in this case.

\paragraph{Indirect effects based on interface exchange coupling}

\begin{figure}
\begin{center}
\includegraphics[width=0.7\textwidth]{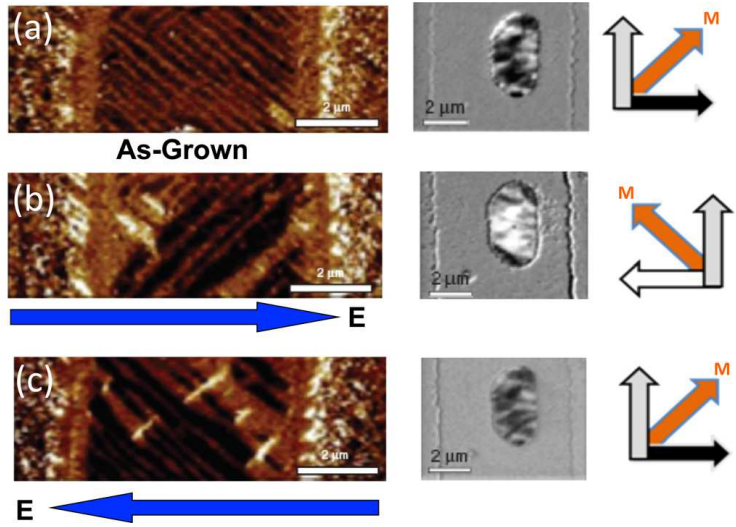}
\caption{BFO/Co$_{0.9}$Fe$_{0.1}$ device allowing magnetization direction control of the soft FM magnetization with an in-plane electric field. Left are PFM images of the BFO domain structure and (middle), XMCD of the soft magnet feature (middle) where an in-plane electric field can be applied (blue arrows). (a) is the as-grown structure while (b) and (c) are the 1$^\mathrm{st}$ and 2$^\mathrm{nd}$ switches. On the right is a schematics illustrating the correlation of XMCD contrast to moment direction and the direction of the total moment (orange arrow). Reprinted from Ref. \cite{Heron2014}.}
\label{fig6}
\end{center}
\end{figure}
 
Intrinsic magnetoelectric coupling in BFO can therefore lead to a change in the direction of AF vectors but possible applications require a global electrically controlled magnetization switching. An appropriate solution can be found using the properties of exchange coupling between a soft ferromagnet (FM) and the antiferromagnetic (AFM) order of BFO. Interface coupling between FM and AFM has been the object of an important research activity in magnetism since its discovery almost 60 years ago \cite{Meiklejohn1956,Meiklejohn1957}. The most obvious experimental signatures associated with magnetic coupling through this interface are a change in the coercivity and a shift or "bias" of the magnetization hysteresis loops of the ferromagnetic layer. An abundant literature has emerged in which various mechanisms for this coupling have been proposed (for reviews, see Refs. \cite{Berkowitz1999,Nogues1999,Stamps2000}). Even though the details of the coupling are not always completely understood, it is clear that this effect represents a convenient way of addressing a net magnetization if one can control the antiferromagnet via the magneto-electric effect. Soft FM layers like CoFeB or NiFe layers can be deposited on BFO thin films \cite{Martin2008,Bea2008a,Dho2009} and single crystals \cite{Lebeugle2009}. A significant interface coupling is found to occur in the two systems, but these differ qualitatively. In thin film based heterostructures on SrTiO$_3$ substrates, both exchange bias and a twofold anisotropy of the FM layer is observed, which does not depend on the particular directions of the BFO antiferromagnetic structure. In contrast, no macroscopic bias is observed in NiFe films deposited on BFO crystals where the anisotropy direction is found to be imposed by the underlying antiferromagnetic structure. There are therefore several important differences between BFO in bulk and thin film forms including the single domain or multidomain state, the cycloidal or collinear G-type AFM structure and the absence or existence of an uncompensated ferromagnetic moment. The last points are of major importance for the exchange bias problem. Indeed, magnetic properties of the FM film in contact are attributed to a net magnetization in the AFM close to the interface due to the presence of uncompensated spins and related to all imperfections including domain walls, atomic steps, interface roughness, and atomic-scale disorder \cite{Antel1999,Ohldag2001,Hase2001,Nowak2002,Sampaio2003}. The uncompensated magnetic moment in the AFM close to the interface is therefore expected to play a central role in establishing a bias field. The comparison of the anisotropy of permalloy Py films on BFO monodomain crystals and of CoFeB or BFO multidomain films \cite{Lebeugle2010} reveals a fundamental difference in the fingerprint of the BFO antiferromagnetic structures. The exchange coupling resulting from these two systems is of a completely different nature. In thin film based heterostructures, the magnetization is pinned in the vicinity of the bias field direction, whereas the anisotropy is driven by the AFM domain structure in single crystals. For multidomain films, the FM magnetic response depends on the rotation of the net magnetization in the BFO out of the anisotropy axis set by field deposition. In single crystal systems, an induced anisotropy in the FM layer along the propagation vector of the cycloid in BFO is observed irrespective of the presence and orientation of a field during deposition. The "bulk" BFO AFM structure is in that case the relevant parameter determining the FM properties thanks to a large effective exchange coupling. Indeed, the AF cycloids imprint their long range structure into a wriggle in the FM layer \cite{Elzo2014} which induces a clear anisotropy axis. For both systems it has been shown that it is possible to locally control ferromagnetism with an electric field using the magneto-electric effect in BFO \cite{Laukhin2006,Chu2008,Lebeugle2009}. In crystals, this is done through a rotation of the interface induced easy axis \cite{Chu2008,Chu2008a}, while in films, the net exchange bias can drive magnetization rotation \cite{Heron2014,Chu2008,Heron2011}. Several devices have been designed and operated on these principles (see the review in Ref. \cite{Heron2014}). It is now possible to electrically switch the soft FM layer's magnetization direction (fig. \ref{fig6}) using in-plane  or out-of-plane  electric fields, thus opening the door of new device functionalities with ultra-low energy consumption. 

\section{Ferroelectric phases in Bismuth-based perovskites}

\subsection{Two families of polar structures}

In this second part, we go beyond the particular case of BFO and address the question of polar phases in Bi-based perovskites in general, motivated by the peculiar chemistry related to the $6s^2$ lone pair electrons of the Bi$^{3+}$. This lone pair is at the origin of a hybridization (leading to directional bonding), a high electronic polarisability ("electronic flexibility"), and polar properties, which might combine with magnetic properties if a magnetic cation, e.g. Fe$^{3+}$ or Mn$^{3+}$, sits on the $B$ site of the perovskite, thus offering an avenue to multiferroicity and interesting magnetoelectric coupling phenomena. 

\addtolength{\hoffset}{-1.5cm}
\begin{table}
\begin{center}
\caption{Structural properties of various Bi- and Pb-based perovskites. The strain $e_\mathrm{tot}$ is Aizu's total strain defined as $\sqrt{\sum e_{ij}^2}$ where the $e_{ij}$ are the components of the spontaneous strain tensor.}
\begin{tabular}{lcc c c c c c c c}
\hline\hline
	& 					& \multicolumn{2}{c}{Polarity} 	&& \multicolumn{2}{c}{Magnetism}	&&	Strain & \\ 
\cline{3-4}\cline{6-7}
														& SG at RT 				& Polarisation							& $T_\mathrm c$	&& $\ \ $Order$\ \ $ & $T_\mathrm N$ or $T_\mathrm c$ &&	$e_\mathrm{tot}$ & Ref.	\\
\hline
BiFeO$_3$										& $R3c$						& 100~$\upmu$C.cm$^{-2}$					& 1098~K		&& AFM 	&	643~K						&& 1.4~\% 	& \cite{Catalan2009}	\\
BiAlO$_3$										& $R3c$						& 10~$\upmu$C.cm$^{-2}$ 					&	820~K			&& --  	& --							&& 1.4~\%		& \cite{Belik2012}		\\
BiScO$_3$										& $C2/c$					& --															& --				&& --		& --							&& 1.9~\%		& \cite{Belik2012}		\\
BiCrO$_3$										& $C2/c$					& --															& --				&& AFM	& 109~K						&& 1.5~\%		& \cite{Belik2012}		\\
BiMnO$_3$										& $C2/c$					& --															& --				&& FM 	& 100~K						&& 1.9~\%		& \cite{Belik2012}		\\
BiCoO$_3$										& $P4mm$					& 120~$\upmu$C.cm$^{-2}$					& $>$600~K	&& AFM	& 470~K						&& 20.2~\%	& \cite{Belik2012}		\\
BiGa$_{0.4}$Fe$_{0.6}$O$_3$	&	$Cm$						& 116~$\upmu$C.cm$^{-2}$					& $>$873~K	&& --		& --							&& 19.2~\%	& \cite{Belik2012a}		\\
BiGa$_{0.7}$Mn$_{0.3}$O$_3$	& $Cm$						& 102~$\upmu$C.cm$^{-2}$					& $>$770~K	&& --		& --							&& 18.8~\%	& \cite{Belik2012a}		\\
BiGa$_{0.4}$Cr$_{0.6}$O$_3$	& $R3c$						& 58~$\upmu$C.cm$^{-2}$						& 850~K			&& --		& --							&& 0.6~\%		& \cite{Belik2012a}		\\
BiNiO$_3$										& $P\overline 1$	& --															& --				&& AFM	& 300~K						&& 1.6~\%		& \cite{Belik2012}		\\
BiInO$_3$										& $Pna2_1$				& No data													& $>$870~K	&& --		& --							&& 4.8~\%		& \cite{Belik2012}		\\
Na$_{1/2}$Bi$_{1/2}$TiO$_3$	& $Cc$						& $\approx$ 38~$\upmu$C.cm$^{-2}$	& 590~K			&& --		& --							&& $\approx$ 1~\%		& \cite{Roedel2009} \\
\hline
PbTiO$_3$										& $P4mm$					& $\approx$ 60~$\upmu$C.cm$^{-2}$	& 760~K			&& --		& --							&& 5.0~\%		& \cite{Lines1977}		\\
PbZrO$_3$										& $Pbam$					& --															& --				&& --		& --							&& 1.0~\%		& \cite{Teslic1998}		\\
PbHfO$_3$										& $Pbam$					& --															& --				&& --		& --							&& 0.8~\%		& \cite{Corker1998}		\\
PbVO$_3$										& $P4mm$					& $>$100~$\upmu$C.cm$^{-2}$				& ?					&& AFM	& $\approx$ 45~K	&& 17.5~\%	& \cite{Oka2008}			\\
PbMnO$_3$										& $P4/mmm$				& --															& --				&& ?		& ?								&& 1.4~\%		& \cite{Oka2009}			\\
\hline\hline
\end{tabular}
\label{tab1}
\end{center}
\end{table}
\addtolength{\hoffset}{1.5cm}

The synthesis of the large Bi-based perovskite family remains a relatively young field, and in many cases even the structure at ambient conditions, not to mention more comprehensive phase diagrams, remains controversial. A recent review by A. Belik \cite{Belik2012} provides a good overview of the state-of-the-art in this field by discussing several significant Bi-based perovskites. In table \ref{tab1}, we compare the room temperature structures observed for different Bi$B$O$_3$ perovskites. For perovskites with magnetic cations, the magnetic order and the transition temperature are reported as well. Similarly, the Curie temperatures and polarization values at ambient conditions are repored when relevant. Some lead-based perovskites are given as well for comparison. 

A first interesting observation is that polar phases make up at most half of the phases reported, if not a minority, so that the presence of Bi$^{3+}$ on the $A$ site is by no means a way to ensure ferroelectricity. This however does not mean that the polar activity of Bi$^{3+}$ is suppressed locally in the non-polar phases. The comparison with lead-based perovskites is useful here: the non-polar PbZrO$_3$ and PbHfO$_3$ are known to be \emph{antiferroelectrics}, a state usually described as an antiparallel arrangement of switchable electric dipoles and characterized by the typical double hysteresis loop. It is plausible that antiferroelectric properties show up in the non-polar Bi-based perovskites as well. In some rare-earth substituted BFO, this has indeed been claimed \cite{Cheng2009,Levin2011}, but experimental evidence is too scarce to allow for more than speculations. 

We now focus on the polar Bi-based perovskites in table \ref{tab1}. They can be discriminated in two families according to the values of their spontaneous strain. By spontaneous strain, we consider the total strain calculated using Aizu's formula \cite{Aizu1970,Salje1993}, taking as a reference the lattice of the ideal cubic perovskite. This enables a quantitative comparison of strains between all structures, even with different space groups. Other strain definitions are possible -- which is why absolute values may appear contradictory in the literature -- but would lead to the same conclusions. 

The first family gathers polar perovskites with "modest" strains ($<2$~\%). It contains notably BFO and the isotructural BiAlO$_3$. Even though BFO is frequently described as a "strongly distorted" perovskite, its total strain appears moderate in comparison with the other members of the family. In this rhombohedral structure, the strains result from both tilts of octahedra and polar cation shifts. The values are very close to conventional ferroelectrics, and they exhibit an elongation along the polarization direction which remains quite moderate ($\approx$ 1--2~\%). BiInO$_3$, for which unfortunately only very few experimental data are available, can be attached to this family as well, although its total strain reaches almost 5~\%, because it does not correspond to a strong elongation along the polarization direction, but is rather fairly isotropic. In a broader sense, we can also include in this family the disordered perovskite Na$_{1/2}$Bi$_{1/2}$TiO$_3$ (NBT), its solid solutions with BaTiO$_3$ (NBT-BT) and the vast collection of compounds derived from them in the search for high-performance, lead-free piezoelectrics. In NBT, the polar activity of bismuth is further complicated by the chemical disorder but can be studied by experimental techniques sensitive to the local environment of the cations. In a neutron pair distribution function study, Keeble et al. have shown how the off-centering of bismuth changes with temperature and how its amplitude contrasts with the very small shifts of the non-polar active Na$^+$ on the $A$-site \cite{Keeble2013}. The relation with the average crystal structure is more complex and out of the scope of this review. 

\begin{figure}
\begin{center}
\includegraphics[width=0.4\textwidth]{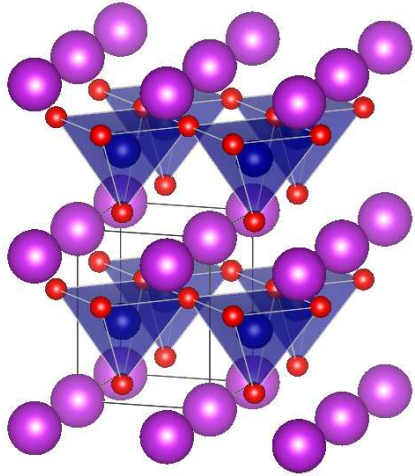}
\caption{Structure of BiCoO$_3$ at 300~K from Ref. \cite{Belik2006}.}
\label{bicoo3}
\end{center}
\end{figure}

The second family of polar Bi-based perovskites gathers compounds that are very strongly distorted, with a total strain of about 20~\%. The distortion corresponds essentially to a very strong elongation along the polar direction. In such phases, The $B$ cation is very strongly shifted away from the center of the octahedron, so that it bonds with five O$^{2-}$ ions instead of six, the last ion being shifted far away in the opposite direction (Fig. \ref{bicoo3}). Such structures are characterized by a very strong polarization, as determined from structural measurements, but also usually higher dielectric losses. BiCoO$_3$ can probably be regarded as the archetype of such structures, with a space group ($P4mm$) only determined by the polar distortion. In the other cases, smaller distortions occur that reduce the symmetry further to $Cm$, but the main structural feature remains the same. It is interesting to mention that such "supertetragonal" phases can also be observed in BFO thin films, and have been in detail investigated by ab-initio calculations \cite{Dieguez2011,Bea2009,Zeches2009}.

Such strongly distorted structures have their equivalent in lead-based perovskites, the archetype being polar PbVO$_3$ \cite{Oka2008,Belik2005a}. It is important to stress that in most cases, polarization switching could not be demonstrated on the bulk material due to large leakage currents and difficulties in sample preparation because of the giant strain \cite{Belik2012a}. Therefore, they can be called "polar" in the strict sense as described e.g. by Kittel, or equivalently "pyroelectric", but not "ferroelectric", the very definition of which implies the possibility to switch the polarization direction. Another consequence is that the experimental values given for the polarization are only derived from structural refinements and point charge models, but not from classical polarization measurements involving polarization reversal. They are therefore compounds that are pretty different from conventional ferroelectrics, and it is not clear whether their specific properties can actually be used in functional devices in the same way as conventional low-strain ferroelectrics. This does not apply to supertetragonal BFO thin films, which can be switched like any ferroelectric film \cite{Bea2009,Yamada2013,Zhang2011a}.

Last, we want to stress that although we have separated the polar phases according to their strain (or elongation), and explained how this strain corresponds to different coordinations of the cations, it is by no means excluded that intermediate cases exist that do not fall easily into these categories. This is best exemplified by the experimental studies of the BiFeO$_3$--BiCoO$_3$ solid solution, which exhibits a continuous transition from the weakly distorted $R3c$ to the strongly distorted $P4mm$ via a monoclinic $Cm$ phase, in a picture that is strongly reminiscent of the monoclinic phases bridging the tetragonal and rhombohedral phases in PZT and similar lead-based solid solution \cite{Oka2012}. Further studies will be needed to elucidate how this huge strain variation of the average structure is accommodated on a local scale. 

In summary, it is clear that the family of Bi-based perovskites display a particularly remarkable diversity of structures, with space groups belonging to nearly every possible crystal systems. We find polar and non-polar phases, and both may combine not only polar or anti-polar shifts of the Bi ion, but also octahedra tilts, sometimes shifts of the $B$ cation, cooperative Jahn-Teller distortion etc. The mere presence of bismuth on the $A$-site of the perovskite is not a sufficient condition for ferroelectricity. Rather, ferroelectricity has to compete with other structural instabilities in the perovskite – notably tilt instabilities, which gives rise to a great variety of structures determined by delicate energy tradeoffs. In this landscape, the "model" compound BFO appears as an exception rather than the rule.

\subsection{Coexistence of ferroelectricity and Jahn-Teller distortion: BiMnO$_3$}

After BFO, the most studied Bi-based perovskite is probably BiMnO$_3$ (BMO). Following the original discussion in N. Hill's paper, its potential multiferroic properties have been widely discussed. The arguments for justifying its ferroelectric properties were the same as those already given for BFO, but magnetic properties are different. The Mn$^{3+}$ cations carry 4$\upmu_B$, against 5$\upmu_B$ for Fe$^{3+}$. BMO exhibits a net ferromagnetic moment below $T_\mathrm N\approx 100$~K. This is original (as seen in Table~\ref{tab1}) since magnetic exchange in most of the oxides is generally due to antiferromagnetic superexchange, as in BFO. It is known that the superexchange depends sensitively on the $B$-O-$B$ angle, which allows considering the existence of an indirect coupling between magnetism and ferroelectricity in such compounds, through a distortive mechanism.  Importantly, the Mn$^{3+}$ ions with a $3d^4:t_{2g}^3e_g^1$ configuration give rise to a strong tendency for Jahn-Teller distortion (JT), while none is expected for Fe$^{3+}$ with $3d^5:t_{2g}^3e_g^2$ in BFO. All this made BMO a potentially interesting and complementary multiferroic. 

Experimental studies, however, have been difficult, mainly because of its demanding synthesis conditions, namely high-pressure synthesis or strain in thin films, and the very structure and ferroelectric character have remained elusive for a long time. Until recently \cite{Wang2009}, BMO was considered to be a "true" multiferroic, combining at low temperature both significant ferroelectricity and ferromagnetism. The combination of electrical polarization hysteresis loops in polycristalline samples, together with reports of a non-centrosymmetric space group $C2$ have fed the idea of ferroelectricity. However, more recently, this viewpoint has been challenged by the report of the non-saturation of electric polarization loops reported, the non-reproducibility of ferroelectric hysteresis curves in bulk samples, thus questioning BMO's ferroelectric nature. Further to this, recent studies of BMO's crystal structure by electron and neutron diffraction of polycrystalline samples do not confirm a $C2$ space group but rather suggest a centrosymmetric space group, $C2/c$, which excludes ferroelectricity. More details on the controversy and the experimental pitfalls are given in \cite{Belik2012}. Ironically, the consensus is now that bulk BiMnO$_3$, the very compound that was taken as a textbook example in N. Hill's seminal paper on polar Bi-based perovskite and in several reviews thereafter \cite{Khomskii2006,Wang2009}, is non polar and therefore not multiferroic.

Nonetheless, the tendency for BMO to develop polar phases does find experimental evidence. In spite of theoretical predictions that did not find any tendency for ferroelectricity under epitaxial strain \cite{Hatt2009}, ferroelectricity has been demonstrated in strained thin films with polarization values of 16--23~$\upmu$C.cm$^{-2}$ \cite{Son2008,Jeen2011,DeLuca2013}, and in La$_{0.1}$Bi$_{0.9}$MnO$_3$ films \cite{Gajek2007}. Another claim was made in \cite{Jung2013}, but the hysteresis loops presented are dominated by dielectric losses and do not prove the ferroelectric character. The role of strain requires deeper clarification though: ferroelectric polarization seems to emerge only for very thin pseudo-cubic films, but not for thicker films where the structure relaxes back to the monoclinic structure \cite{DeLuca2013}. More recently, a polar phase has been suggested to occur in single crystals at very high pressure between 37 and 53~GPa \cite{Guennou2014}. Hysteresis loops cannot be measured at such high pressures, instead the polar character is suggested by a combined analysis of synchrotron XRD and Raman spectroscopy, which leaves only the polar monoclinic space groups $C2$ and $Cm$ as possible candidates. It is also strongly suggested by the huge elongation of the unit cell along one direction: it reaches 8~\% at 53 GPa, which is considerably larger than the distortions seen at lower pressure, larger than the distortion reported for ferroelectric BMO thin film ($c/a = 1.03$, $e_\mathrm{tot} = 2$~\% from Ref. \cite{Son2008}), and even larger than the distortion of the archetypal PbTiO$_3$ at ambient conditions. It admittedly does not reach the strain values of the "supertetragonal" polar phases – which can most probably be accounted for by the considerable compression, but the overview given in table \ref{tab1} shows nonetheless that such a high elongation is only found for polar phases, so that this example too confirms the general idea that a strong elongation of the unit cell is a good indication for ferroelectricity. 

Theoretical approaches will be needed in order to explain the emergence of this polar phase at high pressure in BMO -- and conversely why it does not appear in BFO in a similar pressure range. But some empirical comparison can be made, based on the compounds reported in table \ref{tab1}, on the issue of coexistence between polarity and cooperative JT distortion. Indeed, it is remarkable that all strongly distorted polar phases reported in table~\ref{tab1} have cations on their $B$ site for which JT distortion is expected to be non-existent (Fe$^{3+}$, Cr$^{3+}$, Al$^{3+}$, Ga$^{3+}$) or weak (Co$^{3+}$). Besides, the cooperative long-range JT distortion is strongly reduced under hydrostatic pressure. Another interesting observation is that the solid solution BiMn$_{1-x}$Ga$_x$O$_3$ also shows the $Cm$ polar phase, but only for $x\ge 0.66$, a composition for which the long-range cooperative JT distortion can be expected to be completely suppressed by analogy with the LaMn$_{1-x}$Ga$_x$O$_3$ system \cite{Blasco2002}. Those facts suggest that a strong reduction, if not total suppression, of the cooperative JT distortion, either by hydrostatic pressure of cation substitution, is an important condition for the emergence of this polar phase. This could also be obtained by a combination of pressure and chemical substitution, and such a polar phase should be therefore stable in the $x$–-$P$ phase diagram of BiMn$_{1-x}$Ga$_x$O$_3$ in a much broader region than previously expected. This factor might as well play a role in the stabilization of a ferroelectric phase by epitaxial strain in thin films, but the absence of detailed structural data on thin films does not allow verifying such a conclusion. 

In any case, it is clear that BMO, even non multiferroic in its bulk form, is a fascinating playground for a detailed understanding of the delicate energy competition that govern the structure and properties of Bi-based perovskites. It has a potential to unravel the fundamental solid state chemistry issues to be mastered if Bi-based multiferroics have to find their way towards practical devices.

\end{document}